\documentclass{emulateapj} 
\usepackage{apjfonts}

\begin{document}

\title{High-efficiency photospheric emission of long-duration
  gamma-ray burst jets: the effect of the viewing angle}

\shorttitle{Off-axis GRB photospheric emission}

\author{Davide Lazzati\altaffilmark{1}, Brian
  J. Morsony\altaffilmark{2}, Mitchell
  C. Begelman\altaffilmark{3,4}} \shortauthors{Lazzati, Morsony, \& Begelman}

\altaffiltext{1}{Department of Physics, NC State University, 2401
Stinson Drive, Raleigh, NC 27695-8202}

\altaffiltext{2}{Department of Astronomy, University of
Wisconsin-Madison, 2535 Sterling Hall, 475 N. Charter Street, Madison WI
53706-1582}

\altaffiltext{3}{JILA, University of Colorado, 440 UCB, Boulder, CO
80309-0440}

\altaffiltext{4}{University of Colorado, Department of Astrophysical and
Planetary Sciences, 389 UCB, Boulder, CO 80309-0389}

\begin{abstract} 
  We present the results of a numerical investigation of the spectra
  and light curves of the emission from the photospheres of
  long-duration gamma-ray burst jets. We confirm that the photospheric
  emission has high efficiency and we show that the efficiency
  increases slightly with the off-axis angle. We show that the peak
  frequency of the observed spectrum is proportional to the square
  root of the photosphere's luminosity, in agreement with the Amati
  relation. However, a quantitative comparison reveals that the
  thermal peak frequency is too small for the corresponding total
  luminosity. As a consequence, the radiation must be out of thermal
  equilibrium with the baryons in order to reproduce the
  observations. Finally, we show that the spectrum integrated over the
  emitting surface is virtually indistinguishable from a Planck law,
  and therefore an additional mechanism has to be identified to
  explain the non-thermal behavior of the observed spectra at both
  high and low frequencies.
\end{abstract}


\section{Introduction}

Despite more than thirty years of investigation, the origin of the
prompt emission of GRBs is still veiled in mystery. Among the various
proposed radiation mechanisms are synchrotron emission either in a
baryon- or magnetically dominated jet (Piran 1999; Lloyd \& Petrosian
2000; Zhang \& Yan 2010), synchrotron self-Compton emission (Pe'er \&
Waxman 2004; Baring \& Braby 2004), quasi-thermal comptonization
(Ghisellini \& Celotti 1999), bulk Compton emission (Lazzati et
al. 2000), and photospheric emission, i.e., the radiation advected by
the outflow that is released as the flow becomes optically thin
(Goodman 1986; Rees \& Meszaros 2005; Peer et al. 2005, 2006; Giannios
2006; Lazzati et al. 2009).

In recent years, the photospheric model has gained consensus for its
robust modeling, its lacking of any adjustable parameters, its high
efficiency (Lazzati et al. 2009, hereafter paper I; Mizuta et
al. 2010; Nagakura et al. 2010) and its ability to easily reproduce
the observed peak frequency (Pe'er et al. 2005. 2006; paper
I). Theoretical investigation has moreover revealed that in the case
of sub-photospheric dissipation, photospheric radiation is
characterized by prominent non-thermal high-frequency tails, due to
inverse Compton (IC) scattering of the thermal photons off
relativistic electrons (Pe'er et al. 2006; Giannios 2006; Giannios \&
Spruit 2007; Lazzati \& Begelman 2010). However, due to the relatively
small size of their emission region, photospheric models cannot
explain the very high frequency emission observed by Fermi (Abdo et
al. 2009; Zhang et al. 2010). GeV photons cannot be produced in
photospheric models and their origin has to be identified with the
early afterglow (or external shock) emission (Kumar \& Barniol Duran
2009; Ghisellini et al. 2010). In addition, photospheric models cannot
naturally reproduce the non-thermal low-frequency power-law spectra
below the peak (see, e.g., Pe'er \& Ryde 2010).

In this paper we present an extension of the simulations and results
of Paper I. We have performed a simulation in a 10 times larger box,
extending to $2.5\times10^{13}$~cm, allowing us to explore the
properties of the off-axis emission and to perform a quantitative
comparison of our light curves and spectra with observational data. We
also address the issue of the low-frequency spectrum which is observed
to be non-thermal (Kaneko et al. 2006) but is predicted, in the simple
thermal-photosphere model, to be thermal. In particular we explore
whether the superposition of spectra of different temperatures over the
emitting surface can significantly alter the thermal appearance of
the spectrum (Pe'er \& Ryde 2010).

This paper is organized as follows: in Sect. 2 we describe our
simulations, in Sect. 3 we show the light curves and spectra extracted
from our simulations, in Sect. 4 we compare our synthetic light curves
and spectra to the Amati relation, and in Sect. 5 we discuss our
results in the context of theoretical predictions and simulations from
other groups.

\begin{figure*}
\plotone{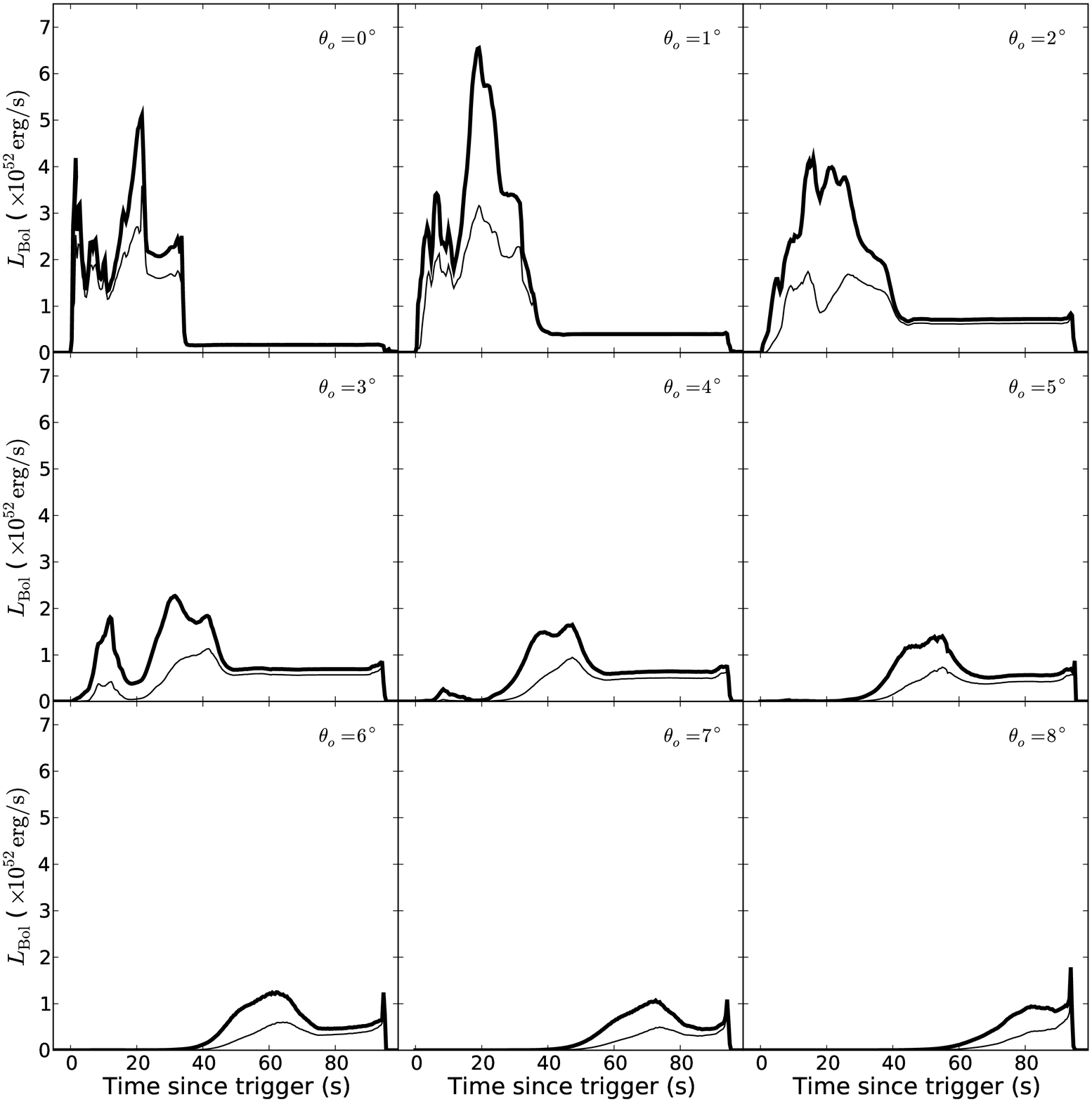}
\caption{{Photospheric light curves for different viewing angles.
    Thick lines show light curves obtained directly from the
    simulations, while thin lines show light curves when a correction
    for the finite size of the simulations box is applied (see text).}
\label{fig:lc}}
\end{figure*}

\section{Numerical Simulations}

The numerical simulation on which this paper is based is an extension
of the one presented in paper I. A jet with luminosity
$L=5.33\times10^{50}$~erg/s is injected in the core of a 16 solar
mass Wolf-Rayet star evolved to pre-explosion (model 16TI, Woosley
\& Heger 2006). The central engine has a constant luminosity for
$100$~s, after which it is sharply turned off. The jet is injected at
a distance of $10^9$~cm from the star center with an opening angle
$\theta_0=10^\circ$, a Lorentz factor $\Gamma_0=5$, and enough
internal energy to reach an asymptotic Lorentz factor
$\Gamma_\infty=400$, upon complete non-dissipative acceleration. The
jet evolution was computed with the special-relativistic, adaptive
mesh refinement code FLASH (Fryxell et al. 2000), as modified by the
authors (Morsony et al. 2007) in a box of size $2.6\times10^{13}$~cm
in the jet direction and $5\times10^{12}$~cm in the equatorial
direction. Science frames were extracted every 0.2 seconds. The
physics included, the resolution, and the refinement schemes were
analogous to those of Morsony et al. (2007) and paper I.

\newpage

\section{Photospheric light curves}

The location of the photosphere in the observer direction (which we
define as the $z$ axis) $Z_{\rm{ph}}$ was calculated analogously to
paper I by back-integrating the optical depth in space and time from a
virtual observer located at a distance of
$Z_{\rm{obs}}=1.3\times10^{13}$~cm. For each time step in the observer
frame, the path of a photon is traced back in time as the photon
travels in the evolving density and velocity patterns. As a result the
location of the photosphere is computed as a function of the observed
time, and no transformation from laboratory to observed coordinates is
required. Note also that we use a Cartesian coordinate $z$ for the
photosphere rather than the radial distance $r$ in order to
automatically take into account the curvature effect and the equal
arriving time surfaces. Let $x$ and $y$ be Cartesian coordinates in
the plane perpendicular to the line of sight and $z$ be the Cartesian
coordinate along the line of sight. We define $x$ to be the coordinate
in the simulation plane and $y$ the coordinate perpendicular to the
simulation plane (and to the line of sight). Given the calculated
comoving density distribution $n^\prime(t_{\rm{lab}},x,z)$, the
location of the photosphere for the photons observed at the time
$t_{\rm{obs}}$ and at a location $x$ is given by:
\begin{equation}
1=-\int_{Z_{\rm{obs}}}^{Z_{\rm{ph}}(x)} 
\sigma_T n^\prime\left(t_{\rm{obs}}-\frac{Z_{\rm{obs}}-z}{c},x,z\right)\,
\Gamma\left[1-\beta\cos(\theta_v)\right]\,dz
\end{equation}
where $\beta\equiv\beta(t_{\rm{lab}},x,z)$ is the local velocity of
the outflow in units of the speed of light,
$\Gamma\equiv\Gamma(t_{\rm{lab}},x,z)$ is the local bulk Lorentz
factor, and $\theta_v\equiv\theta_v(t_{\rm{lab}},x,z)$ is the angle
between the velocity vector and the direction of the line of
sight. All the values of $\beta$, $\Gamma$, and $\theta_v$ are
evaluated at the same delayed coordinate
$(t_{\rm{lab}},x,z)\equiv\left(t_{\rm{obs}}-\frac{Z_{\rm{obs}}-z}{c},x,z\right)$
as the comoving density. Note that when off-axis calculations are
performed (i.e., when the line of sight and the jet axis do not
coincide), the photosphere location becomes a function of $x$ and $y$
and the above equation should be re-written with the explicit
dependence on $y$ as well as on $x$.

Once the photosphere's coordinate $Z_{\rm{ph}}$ is obtained, we
compute the light curves as (paper I\footnote{Note that this equation
  is more precise than the one reported in paper I where the
  integration is erroneously performed over an angular direction
  rather than a linear one. It is also generalized to include lack of
  any axial symmetry, as is the case when the jet axis and the line of
  sight do not coincide.})
\begin{equation}
L(t_{\rm{obs}})=\frac{ac}{2}\int_{-x_{\max}}^{x_{\max}}
\int_{-y_{\max}}^{y_{\max}} dx\,dy
\;\; \frac{T^{\prime 4}}{(1-\beta\cos\theta_v)^2}
\label{eq:lc}
\end{equation}
where $T^\prime=(3p/a)^{1/4}$ is the comoving temperature,
$a=7.56\times10^{-15}$~erg~cm$^{-3}$~K$^{-4}$ is the radiation
constant, and $x_{\max}$ is a distance perpendicular to the line of
sight large enough so that the contribution to the emission at
$x_{\max}$ is negligible. $y_{\max}$ satisfies the same constraint as
$x_{\max}$. Finally, the peak frequency of the observed spectrum is
computed as
\begin{equation}
h\nu_{\rm{peak}}=2.8 \delta k_B T^\prime
\label{eq:hnu}
\end{equation}
were $\delta=1/\Gamma(1-\beta\cos\theta_v)$ is the Doppler factor.

In principle, one would want the observer location $Z_{\rm{obs}}$ to
be approaching infinite, but practical considerations force it to be
inside the computational area. This has the effect of biasing the
photospheric distance, making it smaller than it should be. Assuming
that the evolution at radii larger than the calculated photosphere is
self similar and that there is no further acceleration, one can easily
evaluate the correction thanks to the fact that the opacity in a wind
scales as the inverse square of the distance. Simple algebra yields:
\begin{equation}
\frac{Z_{\rm{ph,true}}}{Z_{\rm{ph,sim}}}=\frac{1}{1-Z_{\rm{ph,sim}}/Z_{\rm{obs}}}
\label{eq:rphcorr}
\end{equation}
where $Z_{\rm{ph,true}}$ is the true location of the photosphere and
$Z_{\rm{ph,sim}}$ is the photospheric radius evaluated from a
simulation with observer location $Z_{\rm{obs}}$. Not surprisingly,
the correction is small for $Z_{\rm{ph,sim}}\ll Z_{\rm{obs}}$, but for
$Z_{\rm{ph,sim}}=Z_{\rm{obs}}/2$ the correction is not
negligible. Under the same assumption of non-accelerating self-similar
expansion, a correction for the light curve and peak frequency can be
obtained, taking into account that the photon-to-baryon ratio does not
change with distance (e.g., Lazzati \& Begelman 2010) and that the
photon temperature scales as $r^{-2/3}$ (e.g., Meszaros \& Rees
2000). The true photospheric luminosity can be computed as:
\begin{equation}
L_{\rm{true}}=L_{\rm{sim}}\left(1-Z_{\rm{ph,sim}}/Z_{\rm{out}}\right)^{2/3}
\label{eq:lccorr}
\end{equation}
and the photon frequency analogously scales as:
\begin{equation}
\nu_{\rm{true}}=\nu_{\rm{sim}}\left(1-Z_{\rm{ph,sim}}/Z_{\rm{out}}\right)^{2/3}
\end{equation}

\begin{figure}
\plotone{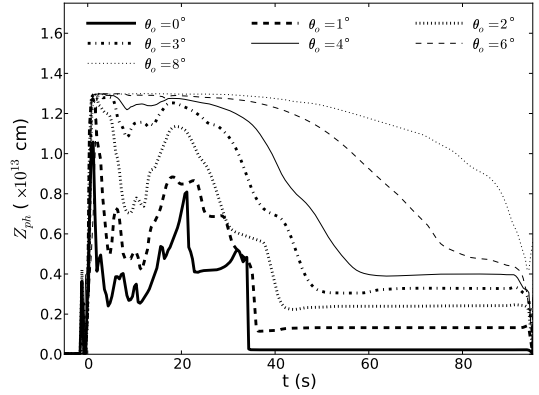}
\caption{{Photospheric distance for the lines of sight analyzed in this
    paper. The photospheric distance is shown for the $(x,y)=(0,0)$
    location, i.e., along the line of sight. Since the distance of the
  virtual observer is $Z_{\rm{obs}}=1.3\times10^{13}$~cm, all the
  photospheric distances that are measured to be close to
  $Z_{\rm{obs}}$ are likely affected by the finite size of the box
  (cfr. Eq.~\ref{eq:rphcorr}). }
\label{fig:rph}}
\end{figure}
 
Figure~\ref{fig:lc} shows the results of the light curve calculation
at nine different observer angles $\theta_o=0$, 1, 2, 3, 4, 5, 6, 7,
and 8 degrees. The light curves at $\theta_o=2.5$ and $3.5^\circ$ were
computed as well but are not shown in the figure. Light curves at
$\theta_o>8^\circ$ were not computed because the photospheric radius
turned out to be too close to the outer boundary of the simulation box
(see Figure~\ref{fig:rph} where all the photospheric radii are
shown). Thick solid lines show the light curves as calculated from the
simulations (Eq.~\ref{eq:lc}) while thin lines show the light curves
once the correction for the finite size of the box is applied
(Eq.~\ref{eq:lccorr}). The comparison between the thick and thin lines
shows that the correction is not large, mostly a factor 2, but becomes
progressively more important at large off-axis angles.  The figure
confirms the results of our paper I (see also Mizuta et al. 2010;
Nagakura et al. 2010) that bright light curves can be produced by
photospheric radiation. Comparing the light curves at different angles
we see that there seem to be a sharp difference between the light
curves at off-axis angles $\theta_o<3$ and those at $\theta_o>3$. The
inner light curves have a bright early phase, characterized by some
level of variability, while the outer light curves have a dim early
phase, becoming more luminous at times when the inner light curves
have lost their brightness. The inner light curves are bright during
the ``shocked jet'' phase, as discussed in Morsony et al. (2007), and
become much more dim in the ``unshocked jet'' phase, during which they
are flat and featureless (at $t\ge40$~s). This difference is due to
the fact that during the ``shocked jet'' phase the jet is
hydrodynamically squeezed and its opening angle is smaller than the
injection opening angle, so that very little emission is seen at
$\theta>\theta_j\sim4^\circ$ (see Figure~6 of Morsony et al. 2010). We
also note that the very sharp rise and decay times seen in the on-axis
light curve are not present in the off-axis ones. We conclude that the
$\theta_o=0^\circ$ light curve is affected by 2D numerical artifacts
and its temporal shape should not be taken as conclusive. We finally
note that the sharp turn-off of the light curves at $t\sim95$~s is due
to the fact that we turned off the engine at $t=100$~s\footnote{The
  $\sim5$ s difference is the time is take the jet to break out of the
  progenitor star.}, and is therefore an artifact of our choice for
the engine duration.

\begin{figure}
\plotone{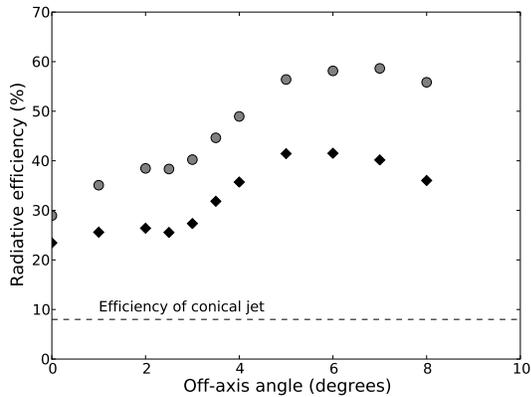}
\caption{{Radiative efficiency of the photosphere's radiation as a
    function of the off-axis observer. Gray symbols show results
    extracted directly form the simulations, while black diamonds show
    the result when the correction for the finite simulation box is
    applied.}
\label{fig:eff}}
\end{figure}
 
An important conclusion we reached in paper I was that of the large
radiative efficiency of the photospheric emission. The radiative
efficiency is computed as the ratio between the energy that is
radiated at the photosphere and the total kinetic energy in the flow
just below the photosphere. In this paper we confirm our previous
finding, even though we point out that the efficiency computation is
severely affected by the finite size of the computation
box. Figure~\ref{fig:eff} shows the efficiency computed directly from
the simulation (gray circles) and the one derived by applying the
finite box correction (Eq.~\ref{eq:lccorr}). The efficiencies computed
in the two ways are different by a factor $\sim2$. In any case, even
the lower estimate puts the efficiency of the photospheric emission in
the $30\%$ range, larger than any efficiency that can be obtained in
internal shocks, without assuming an ad-hoc broad distribution of
Lorentz factors (Lazzati et al. 1999; Panaitescu et al. 1999). The
figure also shows the photospheric efficiency for a conical jet with
the same initial conditions that we adopted in our simulation
($8\%$). Our simulation shows that the jet coupling with the
progenitor star material has the effect of increasing the photospheric
efficiency by a factor 3 to 8, with the highest increase observed at
relatively large off-axis angles (3 to 8 degrees). The efficiency
seems to have a somewhat bimodal behaviour, with the light curves for
$\theta_o<4^\circ$ having a $\sim30\%$ efficiency and the light curves
for $\theta_o>4^\circ$ have a $\sim50\%$ efficiency. The difference of
light curves between small and large angles has been noted above. It
seems that at small angles the light curve is dominated by the
emission from the shocked-jet phase (Morsony et al. 2007), with an
efficiency of $\sim30\%$. At larger off-axis angles, however, the
emission is dominated by the phase during which the jet enters the
corresponding line of sight. In this case the photospheric efficiency
is larger, probably as a cosnsequence of the dissipation due to the
shearing motion of the jet with the progenitor star material.

\begin{figure}
\plotone{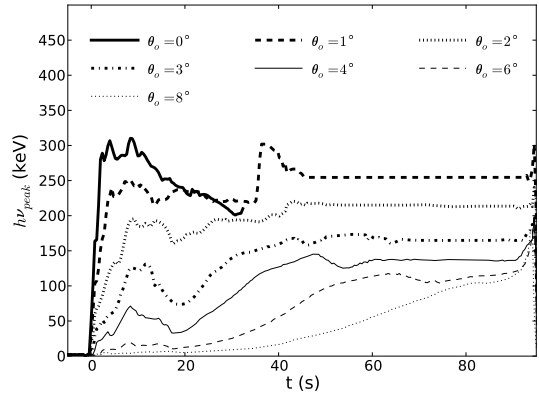}
\caption{{Temporal evolution of the peak frequency of the light curves
    for the various off-axis angles. The $\theta_o=0^\circ$ curve is
    shown only for the first 33 seconds, since a sudden increase to a
    peak frequency of $\sim700$~keV is seen afterwards. This sudden
    increase is due to the onset of the unshocked jet phase (Morsony
    et al. 2007), but the actual value of the peak frequency is a
    numerical artifact due to the coincidence of the line of sight
    with the symmetry axis of the simulation.}
\label{fig:hnu}}
\end{figure}

\begin{figure}
\plotone{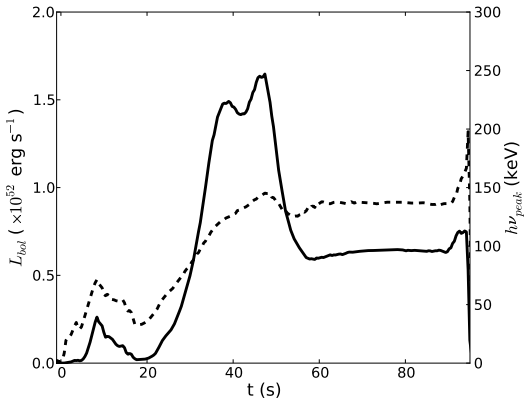}
\caption{{Photospheric light curve for the $\theta_o=4^\circ$ line of
    sight (solid line, left $y$ axis) compared to the evolution of the
    peak frequency of the spectrum (dashed line, right $y$ axis).}
\label{fig:track}}
\end{figure}

Another parameter we extracted from the simulations is the peak
frequency of the light curves (Eq.~\ref{eq:hnu}). Figure~\ref{fig:hnu}
shows the result of the temporal evolution of the peak frequency for
the various lines of sight. First, we note that the observed peak
frequencies are in fairly good agreement with the peak frequencies
observed in GRBs, with an observed mode $h\nu_{\rm{obs}}=250$~keV
(Kaneko et al. 2006). A moderate hard-to-soft evolution is observed
for the $\theta_o=0^\circ$ curve. This is intriguing, given the so far
unexplained hard-to-soft trend in many GRBs. However it could be an
artifact of the $\theta_o=0^\circ$ line of sight, since the trend is
not observed at larger off-axis angles. The comparison of the outer
light curves ($\theta_o>3^\circ$) with their peak frequency evolution
suggests the presence of tracking behaviour, i.e., increasing peak
frequency when the light curve brightens and decreasing peak frequency
when the light curve dims. This behavior is shown in
Figure~\ref{fig:track} for the $\theta_o=4^\circ$ light curve.

A fundamental aspect of the observed GRB spectrum is the presence of
non-thermal power-law tails at either end of the spectrum. The origin
of the high-frequency tail in photospheric emission has been widely
investigated and it has been shown that whenever some level of
dissipation is present in the sub-photospheric region ($10<\tau_T<1$),
inverse Compton processes produce a power-law tail comparable to the
observations, extending up to tens or hundreds of MeV (Pe'er et
al. 2005, 2006; Giannios 2006; Giannios \& Spruit 2007; Lazzati \&
Begelman 2010; Beloborodov 2010). A more serious problem for
photospheric emission models is the presence of non-thermal
low-frequency tails. If the spectrum is generated in a 1-zone
photospheric model with radiation and matter in thermal equilibrium,
non-thermal low-frequency spectra are prohibited since they would
violate the Raleigh-Jeans limit. Pe'er \& Ryde (2010) explored the
possibility of producing low-frequency non-thermal tails by
considering the multi-color black body spectrum integrated over the
emitting surface. They conclude that flat low-frequency spectra
comparable to those observed can be indeed produced by multi-color
photospheres, but only during the decaying part of the light
curve. Pe'er and Ryde (2010) considered emission from conical jets,
for which analytic predictions can be made. We have computed the
multi-color spectrum from our simulations to check whether a
non-thermal spectrum at low-frequencies can be obtained during the
raising part of the light curve, relaxing the conical jet
approximation. The resulting spectrum is shown in Figure~\ref{fig:spex},
compared to a thermal spectrum with the same peak frequency (dashed
line). As the figure shows, we do not find any significant deviation
from a Planck law, because the emission is dominated by the parcel of
the outflow that moves towards the observer. This conclusion is not in
disagreement with the calculation of Pe'er \& Ryde (2010), since they
could find deviations from a thermal spectrum only when the material
pointing directly at the observer is not producing any radiation.

\section{A quantitative comparison}

In order to better evaluate the role of photospheres in the production
of the prompt GRB radiation, we attempt a quantitative comparison of
our results with spectral and luminosity data. It is well known that
observed GRBs for which a redshift measurement is available obey to
the Amati correlation (Amati et al. 2002). According to this
correlation, the burst frame peak frequency is correlated with the
total isotropic equivalent luminosity elevated to an exponent close to
0.5. The most recent best fit to the correlation reads:
\begin{equation}
h\nu_{\rm{peak}} = 102 \left(\frac{E_{\rm{iso}}}{10^{52}
    \rm{erg\,s}^{-1}}\right)^{0.54}
\end{equation}
(Amati, private communication), where the peak frequency is expressed
in keV. The dispersion of the data around the best fit line is
Gaussian with a sigma of $0.2$. Figure~\ref{fig:amati} shows the best
fit correlation with a dashed line and the $3-\sigma$ dispersion areas
with dotted lines. The location of our light curves in the Amati plane
is shown with gray circles (calculated directly from the simulation)
and black diamonds (after the finite box correction). The simulation
points are connected with a line showing the evolution from the
on-axis light curve (upper right point) to the most off-axis light
curve (lowest left point).

The comparison of the simulated light curves with the Amati
correlation is contradictory. On the bright side, as the observer
moves away from the jet axis, the isotropic energy and peak frequency
evolve in such a way as to reproduce the slope of the
correlation. This is independent of whether we consider the corrected
or uncorrected light curves as the most reliable result.  On the other
hand, the normalization is incorrect, since the simulated light curves
are too bright for the corresponding peak frequency (see also Nagakura
et al. 2010).

\section{Discussion}

\begin{figure}
\plotone{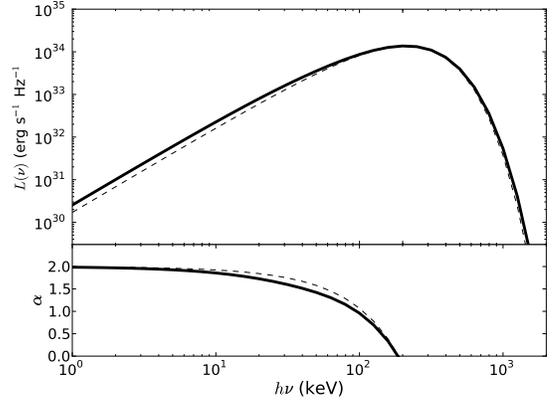}
\caption{{Observed spectrum for the $\theta_o=1^\circ$ observer at
    $t=30$~s, integrated over the emitting surface (upper panel, thick
    solid line). The spectrum shows only small deviations with respect
    to a Planck function with the same maximum (thin dashed line). The
    bottom panel shows the local power-law slope of the two spectra
    ($L(\nu)\propto\nu^{\alpha}$), emphasizing that the difference in
    the slopes is very small.}
\label{fig:spex}}
\end{figure}
 
This paper presents the natural extension of our 2009 numerical study
of the photospheric emission from long-duration GRB jets (paper I). We
here present a simulation run on a domain ten times bigger than in
paper I. Such a simulation allows us to explore the location, spectrum,
and light curves of photospheres along lines of sight different from
the jet axis. The results of this paper substantially confirm the
results of paper I, even though we find that some details of paper I
had been affected by the relatively small computational domain and by
the coincidence of the line of sight with the symmetry axis of the
computation. For example, the sharp features seen in the on-axis light
curve disappear as soon as the observer moves away from the jet
axis. However, we find that the high efficiency of the photospheric
radiation is confirmed and that the efficiency increases for off-axis
lines of sight.

Even with the big simulation box presented here
($Z_{\rm{out}}=2.5\times10^{13}$~cm), the photosphere during some
times interval is sometimes affected by the box size. We therefore
discuss a correction that allows us to derive the correct location of
the photosphere under the approximation of self-similar
non-accelerating expansion beyond the box outer edge. Off-axis
photospheric light curves were discussed elsewhere (Mizuta et al. 2010;
Nagakura et al. 2010). Mizuta et al. perform a simulation with an
identical setup to our paper I. They derive off-axis light curves and
spectra, but their simulation box has an outer edge
$Z_{\rm{out}}=2.5\times10^{12}$~cm, making their results
unreliable. In fact, our Figure~\ref{fig:rph} shows that for all lines
of sight the photosphere is well outside their simulation
box. Nagakura et al. (2010) perform instead a somewhat low-resolution
simulation extending to very large radii. While their simulation is
superior to the one we present here in terms of size, their resolution
and the adoption of a fixed-grid algorithm make their results
inaccurate in terms of the internal properties of the outflow. In
particular, a fixed-grid algorithm is not able to resolve the
tangential shocking that we observe in the jet out to hundreds of
stellar radii, losing the dissipation due to such shocks and therefore
underestimating the temperature at the photosphere.

\begin{figure}
\plotone{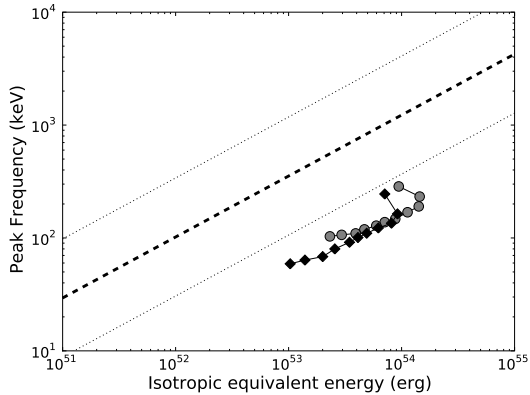}
\caption{{Location of our simulated photospheric light curves and
    spectra in the Amati diagram. The dashed line shows the location of
    the best-fit Amati relation, while the dotted lines show the
    $3-\sigma$ dispersion around it. Black and gray symbols show the
    results with and without the finite-box correction, respectively.}
\label{fig:amati}}
\end{figure}
 
\begin{figure}
\plotone{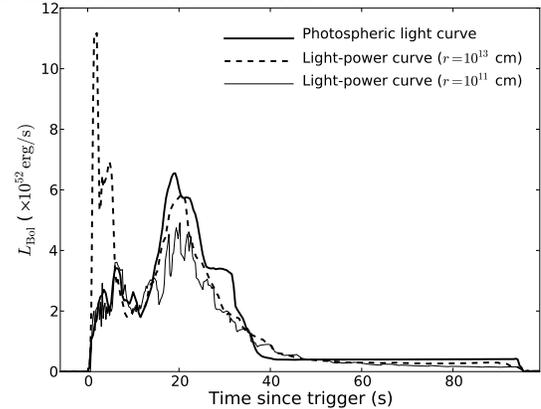}
\caption{{Comparison between the $\theta_o=1^\circ$ photospheric light
    curve (thick solid line) and light-power curves extracted at the
    same viewing angle at radii $r=10^{13}$~cm (thick dashe line) and
    $r=10^{11}$~cm (thin solid line).}
\label{fig:comp}}
\end{figure}

The size of this simulation box allows us to perform an interesting
comparison between the physically motivated photospheric light curves
and the light-power curves that we have used in other publications as
proxies to the real light curve (Lazzati et al. 2010ab; Morsony et
al. 2010). The comparison is shown in Figure~\ref{fig:comp} for the
$\theta_o=1^\circ$ line of sight. Light-power curves were extracted at
radii of $10^{11}$ and $10^{13}$~cm, assuming an efficiency of
$30\%$. We find that in all cases the three bolometric light curves
show a strong correlation, demonstrating that the use of light-power
curves is justified. However, the light-power curve extracted at the
largest radius does show a bright early spike that is not present in
the other curves. This spike is due to the presence of slow, dense
material that piles up ahead of the jet and it may well be an artifact
of the presence of a reflecting boundary condition along the jet
axis. The comparison of photospheric light curves with light-power
curves at different off-axis angles produces results in qualitative
agreement with what shown for the $\theta_o=1^\circ$ case.

Two of our new and more interesting results (not discussed in paper I)
concern the spectrum of the photospheric emission. First, we find that
the integration of the multi-color black-body spectrum over the
emitting surface does not alter substantially the Planck law shape of
the photospheric emission (Figure~\ref{fig:spex}). Second, we find that
the peak frequencies produced by the photosphere are too small for the
isotropic equivalent energy of their respective light curves
(Figure~\ref{fig:amati}). These two results are important because they
show that the photospheric emission has trouble in reproducing the
non-thermal low-frequency spectra observed in GRBs and their peak
frequencies. Both these problems may be solved if the assumption of
thermal equilibrium between the baryons and the radiation field that
we adopted to derive the spectrum are inadequate. As a matter of fact,
the bolometric light curve calculation relies on the simple assumption
that the internal pressure of the outflow is mainly due to radiation
and is therefore quite robust. Such assumption is correct no matter
the spectrum of the radiation field. If, however, the radiation has a
color temperature that is higher than its effective temperature, our
spectral calculations would be incorrect. A first consequence would be
that the peak frequency that we derive is underestimated, and
therefore the Amati correlation discrepancy could disappear. As a
matter of fact, in scattering dominated accretion disk atmospheres, a
discrepacy of a factor $\sim3$ is found in detailed radiation transfer
calculations (e.g., Davis et al. 2005). In addition, having a color
temperature higher than the effective temperature, allows for the
presence of non-thermal low-frequency tails without violating the
Raleigh-Jeans limit.

Even though finding the exact mechanism that would create such an
off-balance configuration between the baryon and the radiation field
is beyond the scope of this paper, we argue that research on the
photospheric emission of GRB jets should proceed in two directions. On
the one hand, it is mandatory to explore the shape of the spectrum of
the photospheric emission from first principles, relaxing the
assumption of equilibrium. On the other hand, the role of different
jet and progenitor properties should be explored, since all the
available simulations have so far concentrated on the progenitor model
16TI (Woosley \& Heger 2006) and on jet configurations analogous to
the one we explored in paper I.

\acknowledgements We thank Kris Beckwith for useful discussions and
Lorenzo Amati for providing us with an updated version of his
correlation.  Resources supporting this work were provided by the NASA
High-End Computing (HEC) Program through the NASA Advanced
Supercomputing (NAS) Division at Ames Research Center. The software
used in this work was in part developed by the DOE-supported ASC /
Alliance Center for Astrophysical Thermonuclear Flashes at the
University of Chicago.This work was supported in part by NASA ATP
grant NNG06GI06G, Fermi GI program NNX10AP55G and Swift GI program
NNX08BA92G.

\end{document}